# Topic Models Conditioned on Arbitrary Features with Dirichlet-multinomial Regression

**David Mimno**
Computer Science Dept.
University of Massachusetts, Amherst
Amherst, MA 01003

**Andrew McCallum**
Computer Science Dept.
University of Massachusetts, Amherst
Amherst, MA 01003

## Abstract

Although fully generative models have been successfully used to model the contents of text documents, they are often awkward to apply to combinations of text data and document metadata. In this paper we propose a Dirichlet-multinomial regression (DMR) topic model that includes a log-linear prior on document-topic distributions that is a function of observed features of the document, such as author, publication venue, references, and dates. We show that by selecting appropriate features, DMR topic models can meet or exceed the performance of several previously published topic models designed for specific data.

## 1  Introduction

Bayesian multinomial mixture models such as latent Dirichlet allocation (LDA) [3] have become a popular method in text analysis due to their simplicity, usefulness in reducing the dimensionality of the data, and ability to produce interpretable and semantically coherent topics. Text data is generally accompanied by metadata, such as authors, publication venues, and dates. Many extensions have been proposed to the basic mixture-of-multinomials topic model to take this data into account. Accounting for such side information results in better topics and the ability to discover associations and patterns, such as learning a topical profile for a given author, or plotting a timeline of the rise and fall of a topic. Currently, developing models for new types of metadata involves specifying a valid generative model and implementing an inference algorithm for that model. In this paper, we propose a new family of topic models based on Dirichlet-multinomial regression (DMR). Rather than generating metadata or estimating topical densities for metadata elements, DMR topic models condition on observed data. As with other conditional models such as maximum entropy classifiers and conditional random fields, users with limited statistical and coding knowledge can quickly specify arbitrarily complicated document features while retaining tractable inference.

The simplest method of incorporating metadata in generative topic models is to generate both the words and the metadata simultaneously given hidden topic variables. In this type of model, each topic has a distribution over words as in the standard model, as well as a distribution over metadata values. Examples of such "downstream" models include the authorship model of Erosheva, Fienberg and Lafferty [5], the Topics over Time (TOT) model of Wang and McCallum [15], the Group-Topic model of Wang, Mohanty and McCallum [16], the CorrLDA model of Blei and Jordan [1] and the named entity models of Newman, Chemudugunta and Smyth [12].

One of the most flexible members of this family is the supervised latent Dirichlet allocation (sLDA) model of Blei and McAuliffe [2]. sLDA generates metadata such as reviewer ratings by learning the parameters of a generalized linear model (GLM) with an appropriate link function and exponential family dispersion function, which are specified by the modeler, for each type of metadata. We show in Section 4.3 that the TOT model is an example of sLDA.

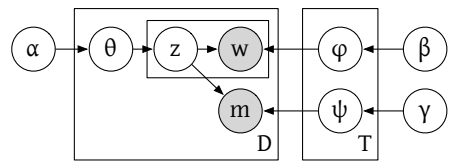

Figure 1: Graphical model representation of a "downstream" topic model, in which metadata $m$ is generated conditioned on the topic assignment variables $z$ of the document and each topic has some parametric distribution over metadata values.

Another approach involves conditioning on metadata elements such as authors by representing document-topic distributions as mixtures of element-specific distributions. One example of this type of model is the author-topic model of Rosen-Zvi, Griffiths, Steyvers and Smyth [13]. In this model, words are generated by first selecting an author uniformly from an observed author list and then selecting a topic from a distribution over topics that is specific to that author. Given a topic, words are selected as before. This model assumes that each word is generated by one and only one author. Similar models, in which a hidden variable selects one of several multinomials over topics, are presented by Mimno and McCallum [11] for modeling expertise by multiple topical mixtures associated with each individual author, by McCallum, Corrada-Emmanuel, and Wang [9] for authors and recipients of email, and by Dietz, Bickel and Scheffer [4] for inferring the influence of individual references on citing papers. These "upstream" models essentially learn an assignment of the words in each document to one of a set of entities.

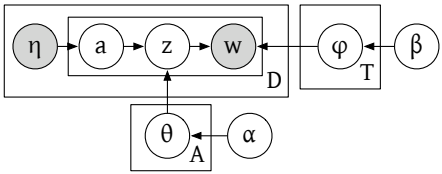

Figure 2: An example of an "upstream" topic model (Author-Topic). The observed authors determine a uniform distribution $\eta$ over authors. Each word is generated by selecting an author, $a$, then selecting a topic from that author's topic distribution $\theta_a$, and finally selecting a word from that topic's word distribution.

Previous work in metadata-rich topic modeling has focused either on specially constructed models that cannot accommodate combinations of modalities of data beyond their original intention, or more complicated models such as exponential family harmoniums and sLDA, whose flexibility comes at the cost of increasingly intractable inference. In contrast to previous methods, Dirichlet-multinomial regression (DMR) topic models are able to incorporate arbitrary types of observed continuous, discrete and categorical features with no additional coding, yet inference remains relatively simple.

In section 4 we compare several topic models designed for specific types of metadata to DMR models conditioned on features that emulate those models. We show that performance of DMR models is in almost all cases comparable to similar generative models, and can be considerably better.

## 2 Modeling the influence of document metadata with Dirichlet-multinomial regression

For each document $d$, let $\boldsymbol{x}_d$ be a vector containing feature that encode metadata values. For example, if the observed features are indicators for the presence of authors, then $\boldsymbol{x}_d$ would include a 1 in the positions for each author listed on document $d$, and a 0 otherwise. In addition, to account for the mean value of each topic, we include an intercept term or "default feature" that is always equal to 1.

For each topic $t$, we also have a vector $\boldsymbol{\lambda}_t$, with length the number of features. Given a feature matrix $X$, the generative process is:

1. For each topic $t$,
    (a) Draw $\boldsymbol{\lambda}_t \sim \mathcal{N}(0, \sigma^2 I)$
    (b) Draw $\boldsymbol{\phi}_t \sim \mathcal{D}(\beta)$

2. For each document $d$,
    (a) For each topic $t$ let $\alpha_{dt} = \exp(\boldsymbol{x}_d^T \boldsymbol{\lambda}_t)$.
    (b) Draw $\boldsymbol{\theta}_d \sim \mathcal{D}(\boldsymbol{\alpha}_d)$.
    (c) For each word $i$,
        i. Draw $z_i \sim \mathcal{M}(\boldsymbol{\theta}_d)$.
        ii. Draw $w_i \sim \mathcal{M}(\boldsymbol{\phi}_{z_i})$.

The model therefore includes three fixed parameters: $\sigma^2$, the variance of the prior on parameter values; $\beta$, the Dirichlet prior on the topic-word distributions; and $|T|$, the number of topics.

Integrating over the multinomials $\boldsymbol{\theta}$, we can construct the complete log likelihood for the portion of the model involving the topics $\boldsymbol{z}$:

$$P(\boldsymbol{z}, \boldsymbol{\lambda}) = \tag{1}$$
$$\prod_d \frac{\Gamma(\sum_t \exp(\boldsymbol{x}_d^T \boldsymbol{\lambda}_t))}{\Gamma(\sum_t \exp(\boldsymbol{x}_d^T \boldsymbol{\lambda}_t) + n_d)} \prod_t \frac{\Gamma(\exp(\boldsymbol{x}_d^T \boldsymbol{\lambda}_t) + n_{t|d})}{\Gamma(\exp(\boldsymbol{x}_d^T \boldsymbol{\lambda}_t))} \times$$
$$\prod_{t,k} \frac{1}{\sqrt{2\pi\sigma^2}} \exp\left(-\frac{\lambda_{tk}^2}{2\sigma^2}\right).$$

The derivative of the log of Equation 1 with respect to the parameter $\lambda_{tk}$ for a given topic $t$ and feature $k$ is

$$\frac{\partial \ell}{\partial \lambda_{tk}} = \tag{2}$$
$$\sum_d x_{dk} \exp(\boldsymbol{x}_d^T \boldsymbol{\lambda}_t) \times$$
$$\Big(\Psi\big(\sum_t \exp(\boldsymbol{x}_d^T \boldsymbol{\lambda}_t)\big) - \Psi\big(\sum_t \exp(\boldsymbol{x}_d^T \boldsymbol{\lambda}_t) + n_d\big) +$$
$$\Psi\big(\exp(\boldsymbol{x}_d^T \boldsymbol{\lambda}_t) + n_{t|d}\big) - \Psi\big(\exp(\boldsymbol{x}_d^T \boldsymbol{\lambda}_t)\big)\Big) - \frac{\lambda_{tk}}{\sigma^2}.$$

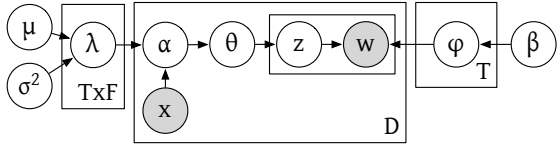

Figure 3: The Dirichlet-multinomial Regression (DMR) topic model. Unlike all previous models, the prior distribution over topics, $\alpha$, is a function of observed document features, and is therefore specific to each distinct combination of metadata feature values.

We train this model using a stochastic EM sampling scheme, in which we alternate between sampling topic assignments from the current prior distribution conditioned on the observed words and features, and numerically optimizing the parameters $\boldsymbol{\lambda}$ given the topic assignments. Our implementation is based on the standard L-BFGS optimizer [8] and Gibbs sampling-based LDA trainer in the Mallet toolkit [10].

## 3 Related Work

Recent work by Blei and McAuliffe [2] on supervised topic models (sLDA) combines a topic model with a log-linear (GLM) model, but in the opposite manner: rather than conditioning on observed features through a log-linear model and then predicting topic variables, sLDA uses topic variables as inputs to the log-linear model to generate observed features. An important advantage of the DMR topic model over sLDA for many applications is that DMR is fully conditional with respect to the observed features. In contrast, sLDA must explicitly estimate probability distributions over all possible feature values by fully specifying the link and dispersion functions for a GLM. Although the class of exponential dispersion families supports a wide range of modalities, the specification of GLMs adds modeling complexity. In addition, adding these distributions to the complete log likelihood of the model may result in a significantly more complicated model that is correspondingly more difficult to train.

In contrast, "off the shelf" DMR topic models can be applied to any set of features with no additional model specification. Furthermore, training a model with complex, multimodal, dependent features is no more difficult in a DMR framework than training a model with a single observed real-valued feature. The distinction between conditional and generative methods is analogous to the difference between maximum entropy and naïve Bayes classifiers and between conditional random fields and hidden Markov models.

Guimaraes and Lindrooth [6] use Dirichlet-multinomial regression in economics applications, but do not use a mixture model or any hidden variables. They observe that Dirichlet-multinomial regression falls within the family of overdispersed generalized linear models (OGLMs), and is equivalent to logistic regression in which the output distribution exhibits extra-multinomial variance. This property is useful because DMR produces unnormalized Dirichlet parameters rather than normalized multinomial parameters. These Dirichlet parameters can then be used as a prior for a Bayesian mixture model.

## 4 Experimental Results

The DMR topic model comprises a broad space of conditional topic models, offering great flexibility for users to define new features. For example, Table 4 shows results for a model incorporating years, venues, authors, and references. In this example, changing two of three authors substantially affects the topical Dirichlet prior. In order to establish that DMR topic models can be effectively trained using the methods described in this paper, we present three examples of DMR models, in which the features are designed to emulate previously published topic models designed for specific types of side information. The purpose of these comparisons is not to suggest that the models compared should necessarily be replaced by equivalent DMR models, but rather to explore the benefits of building custom models relative to simply defining features and passing them to the DMR trainer.

Table 1: DMR topic prior for two documents, given features **2003**, **JMLR**, **D. Blei**, **A. Ng**, and **M. I. Jordan** and features **2004**, **JMLR**, **M.I. Jordan**, **F. Bach**, and **K. Fukumizu**.

| $\alpha_t$ | Topic words (Blei, Ng, Jordan) |
|---|---|
| 2.098 | models model gaussian mixture generative |
| 0.930 | bayesian inference networks network probabilistic |
| 0.692 | classifier classifiers bayes classification naive |
| 0.636 | probabilistic random conditional probabilities fields |
| 0.614 | sampling sample monte carlo chain samples |

| $\alpha_t$ | Topic words (Jordan, Bach, Fukumizu) |
|---|---|
| 4.046 | kernel density kernels data parametric |
| 2.061 | space dimensional high reduction spaces |
| 1.780 | learning machine learn learned reinforcement |
| 1.501 | prediction regression bayes predictions naive |
| 0.879 | problem problems solving solution solutions |

We evaluate the DMR topic model on a corpus of research papers drawn from the Rexa database.[1] For each paper we have text, a publication year, a publication venue, automatically disambiguated author IDs, and automatically disambiguated references. We se-lect a subset of papers from the corpus from venues related to artificial intelligence. We filter out dates earlier than 1987, authors that appear on fewer than five papers, and references to papers with fewer than 10 citations. In addition, for each type of metadata (authors, references, and dates) we train the relevant model only on documents that have that information, since the generative semantics of the Author-Topic model, for example, is undefined if there are no observed authors.

---

[1] http://www.rexa.info

In order to provide a fair comparison and reduce the effect of arbitrary smoothing parameters, we optimize the $\alpha_t$ parameters of all non-DMR topic models using stochastic EM as described by Wallach [14]. This parameter determines the expected mean proportion of each topic. Optimizing the $\alpha_t$ parameters has a substantial positive effect on both model likelihood and held-out performance. Results without hyperparameter optimization are not shown. The DMR model intrinsically represents the mean level of each topic through the parameters for the default feature. The smoothing parameter for the topic-word distributions, $\beta$, is constant for all models at 0.01. The variance $\sigma^2$ for DMR is set to 10.0 for the default features and 0.5 for all other features. All models are run with 100 topics.

We train each model for 1000 iterations. After an initial burn-in period of 250 iterations we optimize parameters ($\lambda$ for DMR, $\alpha$ for all other models) every 50 iterations. All evaluations are run over 10-fold cross validation with five random initializations for each fold.

### 4.1 Author features

For author features, we compare the Author-Topic (AT) model [13] to DMR trained on author indicator features. Example topics for three authors are shown in table 2.

#### 4.1.1 Held-out Likelihood

To evaluate the generalization capability of the model we use the perplexity score described by Rosen-Zvi et al. [13] as well as the empirical likelihood (EL) method advocated for topic model evaluation by Li and McCallum [7]. Evaluating the probability of held-out documents in topic models is difficult because there are an exponential number of possible topic assignments for the words. Both metrics solve this problem by sampling topic distributions from the trained model. Given a trained model, calculating the perplexity score involves sampling topics for half the words in a testing document conditioned on those words. We then use that sampled distribution to calculate the log probability of the remaining words. In the empirical likelihood method, we sample a large number of topic multinomials for each testing document, according to the generative process of the model. We then calculate the log of the average probability of the words given those sampled topic distributions.

Both metrics measure a combination of how good the topic-word distributions are and how well the model can guess which combinations of topics will appear in a given document. The difference is that empirical likelihood estimates the probability of the words without knowing anything about the content of the document, while perplexity also measures the model's ability to "orient" itself quickly given a small amount of local information, such as the first half of the document.

For the EL DMR topic model, we sample $|S|$ unconditional word distributions for a given held-out document $d$ by first calculating the $\alpha_d$ parameters of the Dirichlet prior over topics specific to that document given the observed features $\boldsymbol{x}_d$ in the manner described earlier. We then sample a topic distribution $\boldsymbol{\theta}_{ds}$ from that Dirichlet distribution. Finally, we calculate the probability of each of the observed word tokens $w_i$ by calculating the marginal probability over each topic $t$ of that type using the current point estimates of $P(w_i|t)$ given the topic-word counts.

$$EL(d) = \frac{1}{|S|} \sum_s \sum_i \sum_t \theta_{dts} \frac{n_{w_i|t} + \beta}{n_t + |T|\beta} \quad (3)$$

Results for perplexity and empirical likelihood for AT and DMR with $x_d$ = author indicator functions are shown in Figure 4. DMR shows much better perplexity than either LDA or AT, while both author-aware models do substantially better than LDA in empirical likelihood. The AT models are consistently slightly better in EL than DMR, but the difference is much less than the difference between DMR and LDA. One explanation for the improved perplexity is that DMR uses a "fresh" Dirichlet prior for each held-out document, which can rapidly adapt to local word information in the test documents. In contrast, AT uses multinomials to represent author-topic distributions. These multinomials have less ability to adapt to the test document, as they generally consist of hundreds of previously assigned words.

#### 4.1.2 Predicting Authors

In addition to predicting the words given the authors, we also evaluate the ability of the Author-Topic (AT) and DMR models to predict the authors of a held-out document conditioned on the words. For each model we can define a non-author-specific Dirichlet prior on topics. For AT, defining a prior over topics is equiv-alent to adding a single new, previously unseen author for each held-out document. The Dirichlet prior is specified using the $\alpha$ parameters that are fitted in training the model. For DMR, the topic prior Dirichlet is specified using the prior for a document with no observed features (the exponentiated parameters for the intercept terms for each topic).

Table 2: Ranked topics for three authors under the DMR topic model (left) and the Author-Topic (AT) model (right). For DMR, the sampling distribution for the first word in a document given an author is proportional to the number on the left. For a given topic $t$, this value is $\exp(\lambda_{t0} + \lambda_{ta}x_{da})$, where $\lambda_{t0}$ is the default parameter for topic $t$. For AT, the sampling distribution for the next word $(i+1)$ in a document given the author is proportional to the number on the left. The integer portion generally corresponds to the number of words in a given topic currently assigned to the author, while the fractional part corresponds to $\alpha_t$. These values are much larger than those for DMR, meaning that the topic drawn for word $i+1$ will have relatively little influence on the topic drawn for word $i+2$.

| DMR | | AT | |
|---|---|---|---|
| **David Blei** | | **David Blei** | |
| 0.25 | data mining sets large applications | 48.21 | bayesian data distribution gaussian mixture |
| 0.24 | text documents document categorization large | 36.17 | text documents document information |
| 0.16 | problem work set general information | 31.39 | model models probabilistic modeling show |
| 0.16 | method methods results proposed set | 28.09 | inference approximation propagation approximate |
| 0.15 | distribution bayesian model gaussian models | 15.10 | markov hidden models variables random |
| 0.15 | semantic syntactic lexical sentence named | 11.05 | discourse sentences aspect semantic coherence |
| 0.14 | retrieval information document documents relevance | 9.31 | process approaches methods techniques terms |
| 0.13 | model models show parameters order | 9.20 | probability distribution distributions estimates |
| 0.13 | image images resolution pixels registration | 9.11 | segmentation image texture grouping region |
| 0.11 | translation language word machine english | 8.25 | data sets set large number |
| 0.11 | control robot robots manipulators design | 7.28 | method methods propose proposed applied |
| 0.10 | reasoning logic default semantics theories | 6.15 | networks bayesian probabilistic inference network |
| 0.10 | simple information form show results | 5.28 | problem problems solving solution solutions |
| 0.09 | system systems hybrid intelligent expert | 5.19 | task tasks performed goal perform |
| **Andrew Ng** | | **Andrew Ng** | |
| 0.31 | show algorithms results general problem | 202.11 | reinforcement policy state markov decision |
| 0.31 | number large size small set | 112.30 | error training data parameters sample |
| 0.21 | system systems hybrid intelligent expert | 97.18 | learning bounds function bound algorithms |
| 0.20 | method methods results proposed set | 58.33 | show results problem simple class |
| 0.19 | results quality performance show techniques | 57.36 | algorithm algorithms efficient problem set |
| 0.18 | algorithm algorithms efficient fast show | 54.26 | optimal time results computing number |
| 0.17 | learning reinforcement policy reward state | 39.21 | bayesian data distribution gaussian mixture |
| 0.16 | decision markov processes mdps policy | 34.37 | learning learn machine learned algorithm |
| 0.15 | feature features selection classification extraction | 31.37 | work recent make previous provide |
| 0.15 | results experimental presented experiments proposed | 31.25 | set general properties show defined |
| 0.14 | performance results test experiments good | 31.14 | classification classifier classifiers accuracy class |
| 0.14 | learning training learn learned examples | 31.09 | inference approximation propagation approximate |
| 0.14 | knowledge representation base acquisition bases | 30.19 | feature features selection classification performance |
| 0.13 | problem work set general information | 22.39 | model models probabilistic modeling show |
| **Michael Jordan** | | **Michael Jordan** | |
| 0.69 | distribution bayesian model gaussian models | 58.18 | learning bounds function bound algorithms |
| 0.39 | algorithm algorithms efficient fast show | 57.09 | inference approximation propagation approximate |
| 0.38 | show algorithms results general problem | 33.21 | bayesian data distribution gaussian mixture |
| 0.31 | problem work set general information | 27.39 | model models probabilistic modeling show |
| 0.31 | models model modeling probabilistic generative | 27.20 | probability distribution distributions estimates |
| 0.21 | performance results test experiments good | 27.05 | program programs programming automatic |
| 0.21 | problem problems solving solution optimization | 24.17 | entropy maximum criterion criteria optimization |
| 0.19 | learning training learn learned examples | 22.33 | show results problem simple class |
| 0.15 | networks bayesian inference network belief | 21.25 | set general properties show defined |
| 0.14 | data mining sets large applications | 20.36 | algorithm algorithms efficient problem set |
| 0.12 | simple information form show results | 20.11 | kernel support vector machines kernels |
| 0.12 | function functions gradient approximation linear | 18.14 | methods simple domains current incremental |
| 0.12 | methods techniques approaches existing work | 18.02 | genetic evolutionary evolution ga population |
| 0.12 | learning machine induction rules rule | 17.19 | feature features selection classification performance |

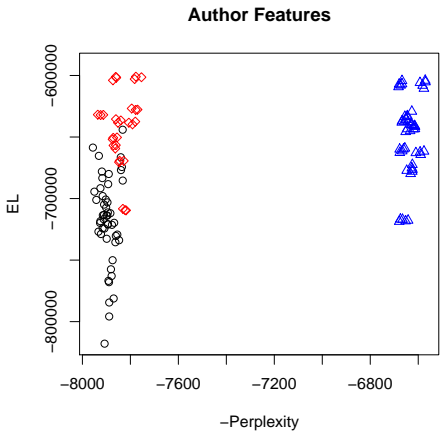

Figure 4: Perplexity and empirical log likelihood for the DMR topic model trained with author indicator features (blue triangles), the Author-Topic (AT) model (red diamonds), and LDA (black circles). Clusters of points represent cross-validation folds. Perplexity is much better for DMR than either AT or LDA. Empirical likelihood is better for the author-aware models than for LDA. In every case, AT performs slightly better in EL than DMR.

For each held-out document, we independently sample 100 sequences of topic assignments from the generative process defined by the model, given the word sequence and the topic prior. We add up the number of times each topic occurs over all the samples to get a vector of topic counts $n_1...n_{|T|}$. We then rank each possible author by the likelihood function of the author given the overall topic counts. For AT, this likelihood is the probability of adding $n_t$ counts to each author's Dirichlet-multinomial distribution, which is defined by the number of times each topic is assigned to an author $n_{t|a}$ and the total number of tokens assigned to that author $n_a$:

$$P(d|a) = \frac{\sum_t \alpha_t + n_a}{\sum_t \alpha_t + n_a + \sum_t n_t} \prod_t \frac{\alpha_t + n_{t|a} + n_t}{\alpha_t + n_{t|a}} \quad (4)$$

For DMR, we define a prior over topics given only a particular author as the Dirichlet parameters under the DMR model for a document with only that author feature; in other words, the exponentiated sum of the default feature parameter and the author feature parameter, for each topic. The likelihood for an author is the Dirichlet-multinomial probability of the $n_t$ counts with those parameters. Note that the likelihoods for a given author and held-out document are not necessarily comparable between DMR and AT, but what we are interested in is the ranking.

Results are shown in Figure 5. DMR ranks authors consistently higher than AT.

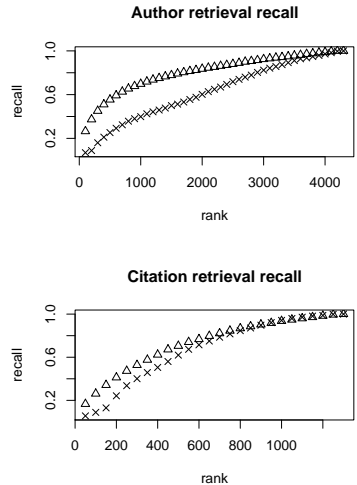

Figure 5: Prediction results for authors and citations. DMR is shown with triangles, and AT and Citation topics with Xes.

## 4.2 Citation features

Following Dietz, Bickel and Scheffer [4], we consider a model for citation influence that is similar to the Author-Topic model. Each citation is treated as a potential "author", such that when the model generates a word, it first selects a paper from its own references section and then samples a topic from that paper's distribution over topics.

Empirical likelihood results for this citation model are, like AT, slightly better than a DMR model with the same information encoded as citation indicator features. Perplexity was significantly better for the citation model. In this case, the number of occurrences of citations may allow the generative model to obtain a better representation of the topical content of citations. In contrast to authors, each citation's topic multinomial may only consist of a few dozen words, allowing it to adapt more easily to local information.

The DMR model also shows citation prediction performance comparable to the generative citation topic model, as shown in Figure 5.

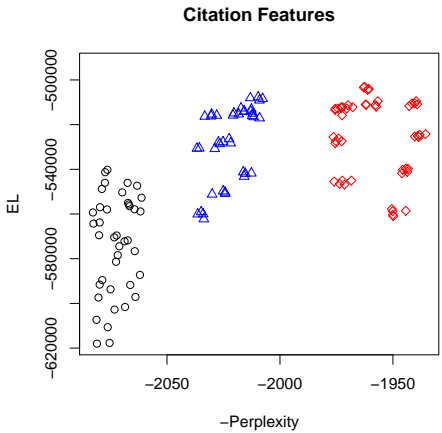

Figure 6: Perplexity and empirical log likelihood for the DMR topic model trained with citation features, the Citation model, and LDA. Unlike other features, citations show very strong perplexity results for the upstream Citation model. DMR continues to outperform LDA in this metric. As with the Author-Topic model, the Citation model has slightly better empirical likelihood than DMR, and both substantially outperform LDA.

### 4.3 Date features

In many text genres, the date of publication provides information about the content of documents. For example, a research paper published in an artificial intelligence conference in 1997 is much more likely to be about neural networks and genetic algorithms than about support vector machines. The opposite is likely to be true of a paper published in 2005.

Previous work on topic models that take into account time includes the Topics over Time (TOT) model of Wang and McCallum [15]. As with LDA, under the TOT model each word $w_i$ is generated by a hidden topic indicator variable $z_i$. In addition, the TOT generative process also samples a "date" variable from a topic-specific beta distribution parameterized by $\psi_{t1}$ and $\psi_{t2}$ $\forall t \in T$. The support of the beta distribution is real numbers between zero and one, so rather than generating an actual date, TOT generates a point proportional to the date of a document, within a finite range of dates. We define this proportion, $p_d = \frac{date_d - \min_{d'} date_{d'}}{\max_{d'} date_{d'} - \min_{d'} date_{d'}}$. In order to sample efficiently, Wang and McCallum use the convention that rather than generating the date once per document, each word in a given document generates its own date, all of which happen to be the same.

Consider the terms in the likelihood function for a TOT model that involve $p_d$ for some document $d$:

$$P(p_d|\mathbf{z}_d) = \qquad (5)$$
$$\prod_i \frac{1}{Z_{z_i}} \exp\left(\psi_{z_i 1} \log(p_d) + \psi_{z_i 2} \log(1 - p_d)\right)$$

where $Z_t$ is the beta function with parameters $\psi_{t1}$ and $\psi_{t2}$. Since $p_d$ is constant for every token in a given document, we can rewrite Equation 6 as

$$\frac{1}{Z} \exp\left(\sum_i \psi_{z_i 1} \log(p_d) + \psi_{z_i 2} \log(1 - p_d)\right) (6)$$

From this representation, we can see two things. First, this expression is the kernel of a beta distribution with parameters $\sum_i \psi_{z_i 1}$ and $\sum_i \psi_{z_i 2}$, so $Z$ is equal to a beta function with those parameters. Second, this expression defines a generalized linear model. The link function is identity, the exponential dispersion function is beta, and the linear predictor is a function of the number of words assigned to each topic, the topic beta parameters, and the sufficient statistics, which are $\log(p)$ and $\log(1-p)$. With the slight modification of substituting normalized topic counts $\bar{z} = 1/N \sum_i z_i$ for the raw topic counts, we see that TOT is precisely a member of the sLDA family of topic models [2].

To compare DMR regression topic models to TOT, we use the same sufficient statistics used by the beta density: $x_d = \log(p_d)$ and $\log(1-p_d)$. DMR and TOT therefore have the same number of parameters: two for each topic date distribution, plus one parameter (the topic intercept parameter in DMR, an optimized $\alpha_t$ for TOT) to account for the mean proportion of each topic in the corpus.

Figure 7 shows perplexity and EL results for TOT and DMR with TOT-like features. DMR provides substantially better perplexity, while also showing improved empirical likelihood.

## 5 Conclusions

The Dirichlet-multinomial regression topic model is a powerful method for rapidly developing topic models that can take into account arbitrary features. It can emulate many previously published models, achieving similar or improved performance with little additional statistical modeling or programming work by the user.

One interesting side effect of using the DMR model is efficiency. Adding additional complexity to a topic model generally results in a larger number of variables to sample and a more complicated sampling distribution. Gibbs sampling performance is mainly a function of the efficiency of the innermost loop of the sampler; in the case of LDA this is the calculation of the sampling distribution over topics for a given word. The Author-Topic model adds an additional set of hidden author assignment variables that must be sampled. TOT adds an additional term (a beta density) to this calculation. In contrast, in a DMR model, all information from the observed document metadata is accounted for in the document-specific Dirichlet parameters. As a result, the sampling phase of DMR training is no more complicated than a simple LDA sampler. The additional overhead of parameter optimization, which we have found decreases as the model converges, can be more than made up by a faster sampling phase, especially if the number of sampling iterations between optimizations is large.

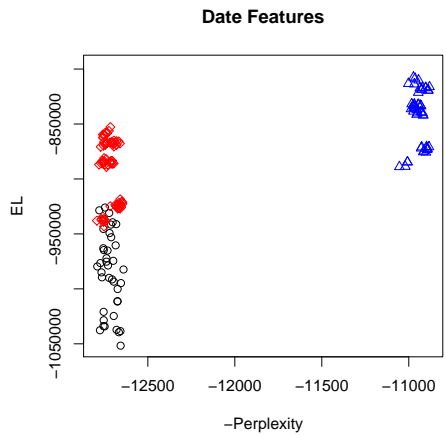

Figure 7: Perplexity and empirical log likelihood for the DMR topic model trained with date features, the Topics Over Time (TOT) model, and LDA. TOT shows perplexity roughly equivalent to LDA, while DMR perplexity is substantially better. DMR also outperforms TOT in empirical likelihood.

DMR provides a useful complement to generative models such as AT and sLDA, which can make inferences about hidden variables and can be incorporated into more complicated hierarchical models. One area for future work is hybrid sLDA-DMR models constructed by splitting the observed features into a set of conditioned variables and a set of generated variables.

## Acknowledgments

Thanks to Xuerui Wang for pointing out the connection between TOT and sLDA, and Hanna Wallach, Rob Hall, and Michael Lavine for helpful discussions.

This work was supported in part by the Center for Intelligent Information Retrieval, in part by NSF grant # CNS-0551597, in part by NSF Nanotech # DMI-0531171, in part by The Central Intelligence Agency, the National Security Agency and National Science Foundation under NSF grant #IIS-0326249, and in part by the Defense Advanced Research Projects Agency (DARPA), through the Department of the Interior, NBC, Acquisition Services Division, under contract number NBCHD030010. Any opinions, findings and conclusions or recommendations expressed in this material are the authors' and do not necessarily reflect those of the sponsor.